\documentclass[aps,prb,reprint]{revtex4-1}

\usepackage{color}
\usepackage{graphicx}
\usepackage{amsmath}
\usepackage{bm}

\let\vec\mathbf

\begin{document}


\title{Type II InAs/GaAsSb Quantum Dots: Highly Tunable Exciton Geometry and Topology}


\author{J. M. Llorens}
\author{L. Wewior}
\author{E. R. Cardozo de Oliveira}
\altaffiliation{On leave from: Departamento de F\'{i}sica, Universidade Federal de S\~{a}o Carlos, 13.565-905, S\~{a}o Carlos, S\~{a}o Paulo, Brazil.}
\affiliation{IMM-Instituto de Microelectr\'{o}nica de Madrid (CNM-CSIC), Isaac Newton 8, PTM, E-28760 Tres Cantos, Madrid, Spain}
\author{J. M. Ulloa}
\author{A. D. Utrilla}
\author{A. Guzm\'{a}n}
\author{A. Hierro}
\affiliation{Institute for Systems based on Optoelectronics and Microtechnology (ISOM), Universidad Polit\'{e}cnica de Madrid, Ciudad Universitaria s/n, 28040 Madrid, Spain}
\author{B. Al\'{e}n}
\email[Corresponding author: ]{benito.alen@csic.es}
\affiliation{IMM-Instituto de Microelectr\'{o}nica de Madrid (CNM-CSIC), Isaac Newton 8, PTM, E-28760 Tres Cantos, Madrid, Spain}


\begin{abstract}
    External control over the electron and hole wavefunctions geometry and
    topology is investigated in a p-i-n diode embedding a dot-in-a-well
    InAs/GaAsSb quantum structure with type II band alignment.  We find highly
    tunable exciton dipole moments and largely decoupled exciton recombination
    and ionization dynamics. We also predict a bias regime where the hole
    wavefunction topology changes continuously from quantum dot-like to quantum
    ring-like as a function of the external bias.
    All these properties have great potential in advanced
    electro-optical applications and in the investigation of fundamental
    spin-orbit phenomena.
\end{abstract}

\pacs{78.55.Cr,78.47.jd,78.67.Hc,73.21.La}

\maketitle


Semiconductor quantum dots (QDs) play a key role in ultrafast signal processing devices such as electro-optical modulators, switches and delay generators.~\cite{Ishikawa_book,SlowLightQDs_2003,Jin2014} Made of different semiconductor materials, like GaAs, InP, Si/Ge,\dots, semiconductor based devices work stable at ultrahigh speeds using a fraction of the energy and space needed by competing electro-optical materials.~\cite{Wada2011} Band-gap engineered semiconductors, and in particular QDs and quantum wells (QW), have made possible this great success thanks to a precise control of electronic excitations via an externally applied electric (or electromagnetic) field; a key concept which, beyond the modulation of light phase and amplitude, underpins many other cutting-edge quantum technologies from excitonic integrated circuits to quantum memories.~\cite{high_control_2008,Gammon_molecules_2006,Krenner2008,Kroutvar2004}

For many years, the prototypical system to develop new classical and quantum
information technologies has been self-assembled In(Ga)As/GaAs QDs. Owing to its type I band alignment, electrons and holes in this system are strongly confined within the In(Ga)As material, being the response of their ground state wavefunctions to the external bias correspondingly small and similar for both carriers.~\cite{fry_inverted_2000,warburton_giant_2002,
finley_quantum-confined_2004} However, for the type II band alignment, in which only the hole or electron is confined within the QD whereas the companion particle remains weakly localized outside, the electro-optical response is comparatively unexplored. It is expected that the weakly
confined particle would be more sensitive to external stimuli,
resulting in a larger electrical polarizability along with other characteristics associated to the very
different confinement regime for electrons and holes.~\cite{Diego_GaSb_2007,Janssens2003} All these features
might vary between both type I and type II QDs and deserve thus a careful
investigation.

\begin{figure}[t]
	\centering
	\includegraphics[width=0.99\columnwidth]{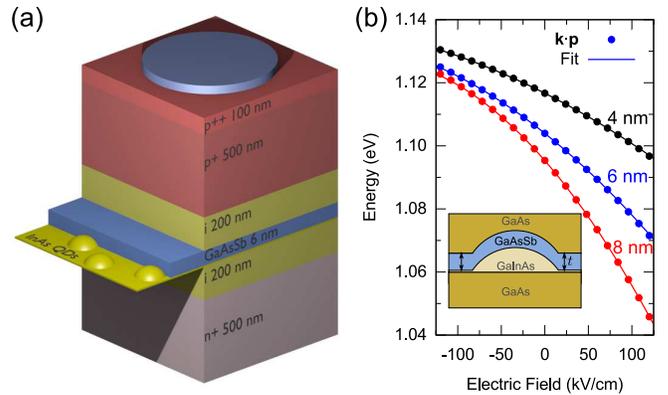}
	\caption{ (a) Schematics of the investigated device structure with
		doping concentrations, layer thicknesses and active layer
		composition indicated. (b) Symbols are k.p calculations of the
		ground state energy dispersion in an external electric field
		for different thicknesses ($t$) of the GaAsSb QW. Black, blue
		and red solid lines are fits to the expression $E(F) = E_0 - p
		F + \beta F^2$ using $p/e$ (nm)= 1.4, 2.2 and 3.2 and $\beta$
		($\mu$eV kV${^{-2}}$cm$^2$)= -0.21, -0.39 and -0.78,
		respectively. The inset shows a detailed scheme of the active layer composition and geometry used for the calculation.
        }
	\label{fig:Sample}
\end{figure}

To explore these fundamental differences, we have embedded an InAs/GaAsSb
dot-in-a-well structure in a GaAs p-i-n diode and investigated its optical
properties as a function of the external electric field at 5 K.
Figure~\ref{fig:Sample}(a) shows an
schematic representation of the device under study. More details about the sample
growth, device fabrication and characterization along with a description of the $8\times 8$
$\vec{k}\cdot\vec{p}$ model that support our findings can be found in the
supplemental material file.~\cite{Note1}

Traditionally, Sb has
been employed either within the InAs QDs~\cite{taboada_structural_2010,taboada_effect_2013} or
in the capping~\cite{ripalda_room_2005,ulloa_gaassb-capped_2010} to shift
their emission wavelength into the O and C optical telecommunications bands. Depending on the amount of Sb, this redshift comes along with a transition
from type I to type II band alignment at the QD interfaces.~\cite{ulloa_analysis_2012,klenovsky_electronic_2010,pavarelli_competitive_2012,hospodkova_type_2013} For the particular
amount of Sb used in the cap layer in the present work, the valence band alignment across the GaAsSb/InAs interfaces
turns from type I to type II, leaving the electrons confined
within the InAs QD, while expelling the holes to the
capping layer already at zero electric field.

Besides the type II band alignment, the vicinity of the top GaAsSb/GaAs interface, located a few nanometers above the QD apex, is a key element in our design (see inset in Figure~\ref{fig:Sample}(b)). Firstly, it acts as an effective barrier for the holes, preventing the weakly localized hole to drift away and ionize from the electron Coulomb potential.~\cite{Janssens2002} Secondly, it allows a precise
control of the key parameters that govern the response to the external field in electro-optical applications. Under these conditions, we find a bias regime where the hole
wavefunction winds around the electron one, inducing a large in-plane dipole
moment. This comes along with a change of the exciton
topology from singly connected to doubly connected, which might have an strong
impact on voltage tunable spin-orbit physics and spintronic devices beyond the
electro-optical realm.~\cite{QR_Book}

The electronic properties of an electron-hole system spatially confined in the
direction of an external electric field, $F$, are governed by the quantum
confined Stark-effect (QCSE).~\cite{Miller1984} Its impact on the ground state energy
can be described up-to second order in perturbation theory by the quadratic function
$E(F) = E_0 - p F + \beta F^2$, where the parameter $p$ represents the
permanent vertical dipole moment and $\beta$ the polarizability.
Figure~\ref{fig:Sample}(b) shows that, despite the complex confinement
geometry, this simple parabolic function accurately describes the ground state
energy evolution calculated using three dimensional eight-band
$\vec{k}\cdot\vec{p}$ methods as implemented in the Nextnano++ software
package~\cite{birner_nextnano:_2007}. It is important to highlight that the
value of $p$ and $\beta$ can be changed only by increasing/reducing the GaAsSb
QW thickness, $t$. This brings a great advantage over type I systems, where
such manipulation relies on changing the dimensions of the QD, which are more
difficult to access experimentally in self-assembled nanostructures.

\begin{figure}[t]
	\centering
	\includegraphics[width=1.00\columnwidth]{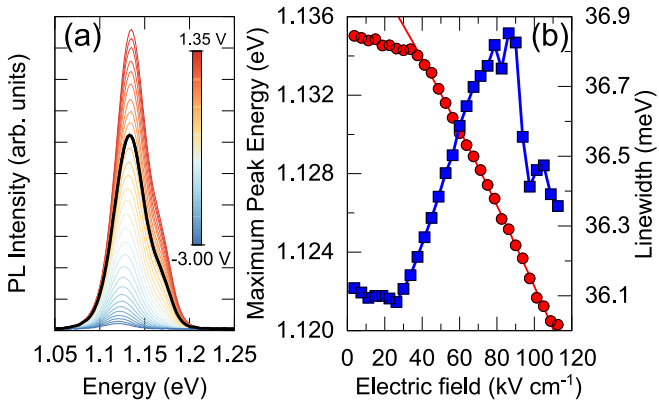}
	\caption{ (a) Evolution of the PL as a function of the external
		voltage. The black thick line corresponds to the 0 V spectrum.
		(b) Ground state peak energy (red dots) and $\mathit{FWHM}$
		(blue squares) extracted from Gaussian deconvolution of the
		main peak. The solid line is the parabolic dependence predicted
		by the $\vec{k}\cdot\vec{p}$ model.
	}
	\label{fig:CW_Optics}
\end{figure}

\begin{figure}[t]
	\centering
	\includegraphics[width=0.80\columnwidth]{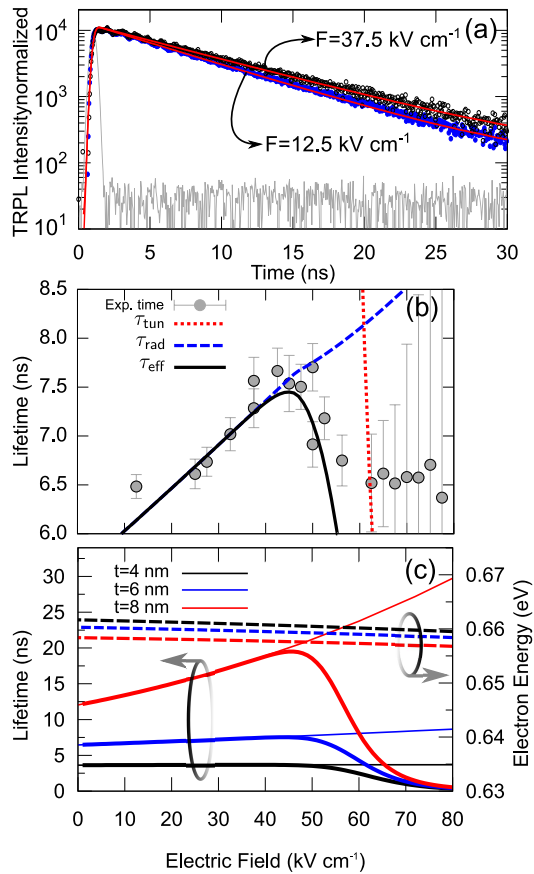}
	\caption{  (a) TRPL experimental data and decay time fits obtained at
		the indicated bias. The system response is plotted in grey. (b)
		Best fit of the theoretical radiative lifetime (dashed line),
		tunneling time (dotted line) and effective lifetime (thick
		solid line) to the experimental decay times (circles). (c) Bias
		evolution of the radiative lifetime (thin solid lines),
		effective lifetime (thick solid lines) and electron energy
		(dashed lines) calculated for different thickness of the GaAsSb
		QW.
	}
	\label{fig:TRPL_Optics}
\end{figure}

Figures~\ref{fig:CW_Optics} and~\ref{fig:TRPL_Optics} shows a direct comparison
between the calculations and experimental data. The evolution of the
photoluminescence spectrum (PL) measured at 5 K as a function of the device bias is displayed
in Figure~\ref{fig:CW_Optics}(a).~\cite{Note2}
At 0 V, the spectrum is dominated by the ground state emission centered at
1.134 eV with full with at half maximum $\mathit{FWHM}\approx36$ meV and a high energy
shoulder related to an excited state transition at 1.168 eV ($\mathit{FWHM}\approx29$
meV). The intensity of the main peak ranges from 79\% of the total emission at
0 V to 94\% at $-3$ V. Therefore, in the following we focus only on the
fundamental transition. Figure~\ref{fig:CW_Optics}(b) shows how from 40 to 110 kV
cm$^{-1}$ the experimental peak energy shifts by 13 meV, nearly a 40\% of the
inhomogeneous bandwidth, while the $\mathit{FWHM}$ remains constant in the same range of electric
fields (variation $<1$ meV).~\cite{Note3}
Both the $\vec{k}\cdot\vec{p}$ method and perturbation theory curves produce a good fit to
the data in this region.  We find a experimental value for $p/e$ = 1.48 nm in
good agreement with the theoretical one $p/e$ = 2.2 nm calculated for
the nominal thickness of the GaAsSb QW, $t$ = 6 nm.
The permanent dipole moment
is related to the expectation value of the electron and hole z coordinate at
$F=$0 kV cm$^{-1}$ through
$p/e=\left<z_\text{e-h}\right>$=$\left<z_\text{h}\right>$-$\left<z_\text{e}\right>$.
The Coulomb interaction is expected to be small due to the type II alignment, but even without it,
our model predicts that the strain and piezoelectric
potentials are enough to stabilize the hole wavefunction above the electron one
at zero bias. In addition, we find that the polarizability $\beta_\text{exp} =
-0.32$ $\mu$eV kV${^{-2}}$cm$^2$ also agrees well with the
theoretical value $\beta_\text{theo} = -0.39$ $\mu$eV kV${^{-2}}$cm$^2$,
concluding that the key parameters of the electrical response of these QDs are
well reproduced by our model.

To analyze the carrier dynamics, we have performed time resolved
photoluminescence experiments (TRPL). Figure~\ref{fig:TRPL_Optics}(a) contains two representative decay
curves recorded at 5 K for two values of the electric field. The solid lines represent
the convolution of the system response with single exponential decay fits for
each dataset. We obtain long decay times, in excess of 6 ns, confirming that
carriers exhibit type II confinement after the rapid thermal annealing treatment.~\cite{ulloa_high_2012,Liu2013} The
full evolution of the decay time with the external electric field is
represented in Figure~\ref{fig:TRPL_Optics}(b). For each point, the decay time and
its statistical error were estimated at the maximum of the emission, i.e. following
the Stark shift of the ground state. The decay time follows a curve typical of
two counteracting processes.~\cite{Alen_oscillator_2007} The region of lifetime increase
($F$ $\lesssim$ 50 kV cm$^{-1}$) corresponds to the vertical separation of the electron
and hole wavefunctions and the reduction of the electron-hole overlap dictated
by the QCSE. The $\vec{k}\cdot\vec{p}$ predicts a
change in the radiative lifetime $\tau_\text{rad}$ from 5 to 12 ns between -150 kV cm$^{-1}$ and
150 kV cm$^{-1}$. The region of
lifetime decrease ($F$ $\gtrsim$ 50 kV cm$^{-1}$) shall be associated to the tunnelling of the electron through
the InAs/GaAs interface. The tunneling rate $\tau_\text{tun}^{-1}$ can be estimated using the
Wentzel-Kramers-Brillouin approximation for a triangular
well:
\begin{equation}{\label{eq1}}
	\frac{1}{\tau_\text{tun}}=\frac{\hbar}{8m^*_eH^2}\text{exp}\left\{-\frac{4}{3}\frac{\sqrt{2m^*_e}}{e\hbar
F}[E_c-E_e(F)]^{3/2}\right\},
\end{equation}
where we introduce the electron effective mass of compressed InAs, $m^*_e=0.1$,
the GaAs band edge at the QD base, $E_c = 0.765$ eV, $H=2.3~$nm, for the QD
height determined independently,~\cite{ulloa_high_2012} and $E_e(F)$ for the
electron ground state energy given by the model. The resulting $\tau_\text{tun}$
is depicted by the dotted line in Figure~\ref{fig:TRPL_Optics}(b). The effective
lifetime is obtained as a sum of the rates $\tau^{-1}$ = $\tau_\text{tun}^{-1}$ +
$\tau_\text{rad}^{-1}$. As displayed by the black solid
curve in Figure~\ref{fig:TRPL_Optics}(b), there is an optimum match with the experimental values
by just rigidly downshifting the theoretical $\tau_\text{rad}$ (dashed line)
by 1 ns.

Large $p$ and $\beta$ are desirable to fabricate electro-absorption
modulators with small modulation voltages and low insertion losses. Due to the
larger polarizability of our QD structure, we expect a noticeable modulation of
the oscillator strength and exciton radiative lifetime with
the applied bias as just described. However, the tunneling of the electron out of the QD prevents
to extend the tuning to higher electric fields.
In this context, an additional feature arises from the large asymmetry between electron and hole
confinement, which allows to tune almost independently $\tau_\text{rad}$ and $\tau_\text{tun}$ in our system.
Figure~\ref{fig:TRPL_Optics}(c) shows how
$\tau_\text{rad}$ and the effective lifetime scales up with the the GaAsSb QW
thickness and applied bias. This is a consequence of the large $\beta$ value of these QDs which is mostly due
to the spatial modulation of the hole wavefunction in the GaAsSb layer. Meanwhile, the strongly confined
electron almost does not change its energy when varying $F$ as depicted by
dashed lines in the same figure (less than 1 meV for a given $t$). At
the same time, due to their deeper confinement potential and large effective
mass, holes do not tunnel through the GaAs/GaAsSb QW interface.  Under this configuration, the tunneling of the electron through the InAs/GaAs interface
determines entirely the exciton ionization and hardly depends
on the
particular GaAsSb layer thickness (only through small variations of the
electron energy due to the strain field). Thus, both parameters can be tuned independently in a wide
range, playing with the GaAsSb QW thickness to fix $\tau_\text{rad}$ and independently introducing lattice matched electron tunneling barriers to fix $\tau_\text{tun}$ as discussed in Ref.~\onlinecite{Giant_shields_2010}.

\begin{figure}[bt]
	\centering
	\includegraphics[width=1.00\columnwidth]{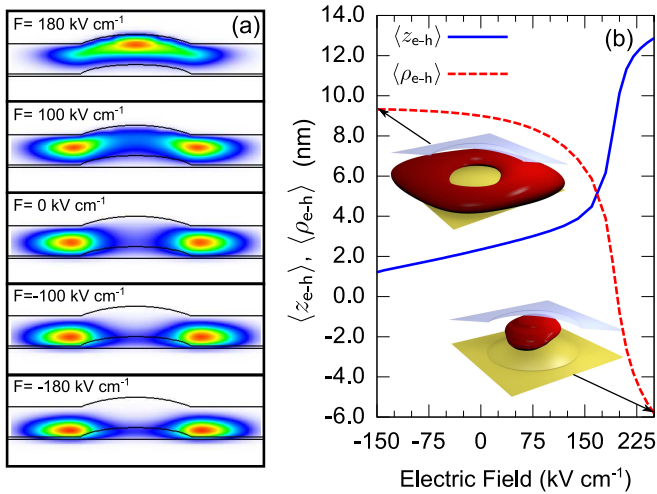}
	\caption{(a) Probability density distribution of the fundamental state
		of holes averaged over (110) and (1$\overline{1}$0) planes. (b)
	Expected value of vertical (blue solid line) and radial (red dashed
line) separation between electron and hole. The insets show three dimensional
colour plots of the ground state hole wavefunction probability density
calculated at the indicated electric fields.}
	\label{fig:Tresults}
\end{figure}

The QCSE properties of the proposed structure do not arise only from a modulation of
the electron hole wavefunction overlap but also from a change of the hole wavefunction topology. An
intuitive picture can be obtained from the probability density calculated by the eight-band $\vec{k}\cdot\vec{p}$ model as shown
in Figure~\ref{fig:Tresults}(a). For $F<$0 kV cm$^{-1}$, the combination of
type II confinement, strain and piezoelectric fields results in a hole
wavefunction that adopts a non-single connected geometry (ring-like) and the
formation of a large in-plane permanent dipole $\left<\rho_\text{e-h}\right>$. Applying a magnetic field in the
growth direction, both conditions shall induce a relative Berry phase between the electron and hole wavefunctions resulting in periodic changes of certain optical magnitudes, like the oscillator strength, and other effects predicted by Aharonov and
Bohm (AB).~\cite{aharonov_significance_1959, Govorov_AB_2002}

These phenomena have received much attention in type I and type II semiconductor nanostructures connected to spin-orbit physics and applications.~\cite{QR_Book} Type-I quantum rings (QRs) with the correct confinement topology for electrons and holes are difficult to fabricate and result in a rather small $\left<\rho_{e-h}\right>$  (typically $\approx$2 nm), making the detection of optical AB effects a very challenging task.~\cite{Govorov_AB_2002,Marcio_AB_2010} The application of a vertical electric field was proposed then to increase slightly $\left<\rho_{e-h}\right>$ and identify tunable AB oscillations in GaAs/AlGaAs QRs.~\cite{ding_gate_2010,li_tunable_2011} To this respect, QDs with type II band alignment might be a better alternative to investigate optical AB phenomena,~\cite{ribeiro_aharonov-bohm_2004, Kuskovsky2007} although the possibility of a bias dependent topology has not been exploited in these systems yet.

Figure~\ref{fig:Tresults}(b) shows the bias evolution of the vertical and in-plane dipole moments
calculated for the In(Ga)As/GaAsSb structure presented here. In contrast to type I systems, at $F=0$ kV cm$^{-1}$, $\left<\rho_\text{e-h}\right>$ is
already 9 nm with $\left<z_\text{e-h}\right>=$2.3 nm. Applying negative
electric fields, the hole moves downwards approaching the electron and
increasing the e-h overlap, as explained before. The in-plane dipole moment
remains mostly unchanged in this bias region which would be the most favorable
for the observation of the optical AB effect. Reversing the field direction,
the hole wavefunction starts to move upwards decreasing $\left<\rho_\text{e-h}\right>$ down
to zero at $F=200$ kV cm${^{-1}}$. A sweep of the lateral dipole moment of this magnitude is unforeseen in the literature and shall pave the way for further
investigations of tunable optical AB effects in semiconductor nanostructures.

In summary, we study the voltage control of the hole wavefunction geometry and topology
in a type-II InAs/GaAsSb dot-in-well structure. We
demonstrate that the key parameters for semiconductor
electro-optical applications in this structure can be tuned precisely in a wider range than using
type-I QDs. We also predict a bias regime where the lateral exciton
dipole moment and the hole wavefunction topology enable the study
of spin-orbit phenomena like voltage tunable optical Aharonov-Bohm effects.

\begin{acknowledgments}

The authors acknowledge the financial support from the Spanish MINECO through the
grants no. ENE2012-37804-C02-02, TEC2011-29120-C05-04 and MAT2013-47102-C2-2-R.
Comunidad de Madrid through the grant no. S2013/MAE-2780. We also thank SGAI-CSIC for the
allocation of computational resources and Brazilian CNPq
for supporting E.R.C. de Oliveira through the grant 313812/2013-6.

\end{acknowledgments}

%

\end{document}